\documentclass[11pt]{article}
\usepackage{latexsym,bm,graphicx,color,xcolor,nicefrac,titletoc,enumerate,amsmath,amssymb,xfrac,xcolor,physics,cite,setspace}

\usepackage{mlmodern}
\usepackage[T1]{fontenc}

\usepackage[nottoc]{tocbibind} 

\usepackage{hyperref}
\hypersetup{linktocpage=true,colorlinks=true,linkcolor=blue,citecolor=blue,urlcolor=blue}

\usepackage[skip=3pt plus1pt, indent=20pt]{parskip}

\usepackage[bottom]{footmisc}

\usepackage{geometry}
\usepackage{titlesec}
\titleformat{\section}{\large\bfseries\sffamily}{\thesection}{0.5em}{}
\titleformat{\subsection}{\normalfont\bfseries\sffamily}{\thesubsection}{0.5em}{}
\titleformat{\subsubsection}{\normalsize\itshape\sffamily}{\thesubsubsection}{0.5em}{}
\titleformat*{\paragraph}{\normalsize\bfseries\sffamily}

\numberwithin{equation}{section}

\def\a{\alpha}
\def\b{\beta}

\def\d{\delta}
\def\D{\Delta}

\def\f{\phi}

\def\m{\mu}
\def\n{\nu}
\def\r{\rho}
\def\s{\sigma}

\def\t{\tau}
\def\th{\theta}

\def\pd{\partial}
\def\nab{\nabla}
\def\pr{\prime}

\def\nn{\nonumber}
\def\qq{\quad\quad}

\newcommand{\sq}{\sqrt}
\newcommand{\sqdet}{\sq{-g}}

\newcommand{\finf}{f_\infty}
\newcommand{\rp}{r_+}

\newcommand{\cC}{\mathcal{C}}

\newcommand{\cF}{\mathcal{F}}
\newcommand{\cG}{\mathcal{G}}

\newcommand{\cL}{\mathcal{L}}

\newcommand{\cN}{\mathcal{N}}
\newcommand{\cR}{\mathcal{R}}

\newcommand{\tD}{\widetilde{\D}}
\newcommand{\tL}{\widetilde{L}}

\newcommand{\rmin}{r_\text{min}}
\newcommand{\onsh}{\eval_{\text{on-shell}}}
\newcommand{\IE}{I_{\text{E}}}

\newcommand{\SC}{S_{\text{Cardy}}}

\newcommand{\mail}[1]{\href{mailto:#1}{{\tt #1}}}

\begin{document}
	
\begin{titlepage}       
	\vspace{5pt} \hfill 
		
		
	\begin{center}
		{\Large \bf \sffamily Microscopic entropy of static black holes in 3d Lovelock gravities}
	\end{center}
		
	\begin{center}
		\vspace{10pt}
			
		{{\bf \sffamily G{\"o}khan Alka\c{c},}${}^{a}\,${\bf \sffamily  Luis Guajardo}${}^{b}\,$ {\bf \sffamily and Hikmet \"{O}z\c{s}ahin}${}^{c}$}
		\\[4mm]
			
		{\small 
		{\it ${}^a$Department of Aerospace Engineering, Faculty of Engineering,\\ At{\i}l{\i}m University, 06836 Ankara, T\"{u}rkiye}\\[2mm]
				
		{\it ${}^b$Instituto de Investigaci\'on Interdisciplinario, Vicerrector\'ia Acad\'emica,\\ Universidad de Talca, 3465548 Talca, Chile}\\[2mm]
		
		{\it ${}^c$Department of Physics, Faculty of Arts and Sciences,\\ Sinop University, 57000 Sinop, T\"{u}rkiye}\\[2mm]
				
		{\it E-mail:} {\mail{alkac@mail.com}, \mail{luis.guajardo.r@gmail.com}, \mail{h.ozsahin@sinop.edu.tr}}
		}
		\vspace{2mm}
		\end{center}
		
		\centerline{{\bf \sffamily Abstract}}
		\vspace*{1mm}
		\noindent We give a microscopic derivation of the semi-classical entropy of static black holes in 3d Lovelock gravities, which are certain 3d Horndeski theories that were recently discovered from higher-dimensional Lovelock gravities via various methods and admit black hole solutions analogous to higher-dimensional ones. Assuming the ground state is described by the soliton obtained from the black hole solution by performing a double Wick rotation, we reproduce the semi-classical entropy from Cardy formula without central charges for the asymptotic growth of the number of states. We explain in detail how the mass of the soliton, which is needed in the Cardy formula, can be computed in the mini-superspace formalism. 
		 
		\vspace{3mm}
		\par\noindent\rule{\textwidth}{0.5pt}
		\tableofcontents
		\par\noindent\rule{\textwidth}{0.5pt}
		\newpage
		\pagestyle{empty}
\end{titlepage}


\section{Introduction} 
One of the most important challanges for a quantum theory of gravity is to provide a microscopic derivation of black hole entropy. Among succesful examples, Strominger's derivation in \cite{Strominger:1997eq} is arguably the simplest since it does not require string theory or supersymmetry. It is based on the observation of Brown and Hanneaux \cite{Brown:1986nw} that the global charges corresponding to asymptotic symmetries of 3d general relativity (GR) with a negative cosmological constant yield a Poisson bracket algebra with a central extension, which turn out to be two copies of the Virasoro algebra with the central charges
\begin{equation}
	c^\pm = \frac{3 \ell}{2 G},\label{cGR}
\end{equation}
where $\ell$ is the AdS radius and $G$ is Newton's constant. Since this is the algebra satisfied by the generators of conformal transformations on the Hilbert space of 2d conformal field theories (CFTs), a consistent quantum theory of gravity on AdS$_3$ can be assumed to be a 2d CFT. The entropy of a 2d CFT can be obtained by means of Cardy formula for the asymptotic growth of the number of states, whose standard form reads \cite{Cardy1986}
\begin{equation}
	\SC = 2 \pi \sqrt{\frac{c^+}{6}  \tD^+} + 2 \pi \sqrt{\frac{c^-}{6} \tD^-},\label{Cardystan}
\end{equation}
where $\tD^\pm$ are the eigenvalues of the shifted Virasoro operators
\begin{equation}
	\tL_0^\pm = L_0^\pm - \frac{c^\pm}{24}.\label{shvir}
\end{equation}
By assuming the additive constants in $\tD^\pm$ vanish for the zero mass and angular momentum black hole ($M=J=0$)  such that
\begin{equation}
	\tD^\pm = \frac{1}{2} (M \ell \pm J),
\end{equation}
Strominger showed that one can reproduce the semi-classical entropy of the Ba\~nados-Teitelboim-Zanelli (BTZ) black hole \cite{Banados1992,Banados1993} from Cardy formula.

Let us see how this works for the static BTZ black hole whose line element is given by
\begin{equation}
	\dd{s}^2_b = -\frac{r^2 - r^2_+}{\ell^2} \dd{t}^2_b + \frac{\ell^2}{r^2-r^2_+} \dd{r}^2 + r^2 \dd{\th}^2_b,\label{BTZgr}
\end{equation}
where $r_+$ is the event horizon radius. This is a useful exercise since we will be interested in static black hole solutions in this paper. By Euclidean methods of \cite{Gibbons1977}, it is easy to get the inverse temperature $\b$, the mass $M$ and the entropy $S$ of the black hole as follows
\begin{equation}
	\b = \frac{2 \pi \ell^2}{r_+}, \qquad \qquad M = \frac{r_+^2}{8 G \ell^2}, \qquad \qquad S = \frac{A}{4 G} = \frac{\pi r_+}{2G},\label{BTZth}
\end{equation}
satisfying the first law $\d S = \b \d M$. Using the value of the central charge given in \eqref{cGR} and $\tD^\pm = \frac{1}{2} M \ell$ (since the angular momentum vanishes, i.e, $J=0$), one sees that Cardy formula \eqref{Cardystan} precisely reproduces the semi-classical entropy given in \eqref{BTZth}, i.e., $\SC = S$.

Later, an intriguing puzzle emerged in \cite{Correa2011} from the study of a static hairy black hole solution that is obtained from minimal coupling of a scalar field to AdS$_3$ gravity with a particular potential. Although the generators of the asymptotic symmetries acquire a nontrivial contribution from the scalar coupling, the Poisson bracket algebra of the global charges remains the same as that of GR with the central charges in \eqref{cGR} \cite{Henneaux:2002wm}. However, when the entropy of the hairy black hole solution is expressed as a function of its mass, it differs from that of the BTZ black hole. Therefore, it seems impossible to reproduce the semi-classical entropy from the Cardy formula \eqref{Cardystan}.

As shown in  \cite{Correa2011}, this issue can be handled by a careful consideration of the Cardy formula from scratch (see \cite{Carlip:1998qw} for a pedagogical derivation). When the lowest eigenvalues $\D^\pm_0$ of the Virasoro operators $L^\pm_0$ are non-vanishing ($\D^\pm_0 \neq 0$), it is given by 
\begin{equation}
	\SC = 2 \pi \sqrt{\frac{c_\text{eff}^+}{6} \tD^+} + 2 \pi \sqrt{\frac{c_\text{eff}^-}{6} \tD^-},\label{Cardyimp}
\end{equation}
where $c_\text{eff}^\pm$ are the effective central charges given by
\begin{equation}
	c_\text{eff}^\pm = c^\pm - 24 \tD^\pm_0.
\end{equation}
This allows us to rewrite the Cardy formula without any reference to central charges $c^\pm$ as
\begin{equation}
	\SC = 4 \pi \sqrt{-\tD^+_0 \tD^+} + 4 \pi \sqrt{-\tD^-_0 \tD^-},\label{Cardynew}
\end{equation}
where $\tD^\pm_0$ are the lowest eigenvalues of the shifted Virasoro operators $\tL^\pm_0$ that we defined in \eqref{shvir}. Once we assume that the BTZ black hole and the hairy black hole belong to disconnected sectors with distinct ground states, the Cardy formula in \eqref{Cardynew} might reproduce the semi-classical entropy. In \cite{Correa2011}, this was achieved by assuming the ground state of each sector is described by the soliton solution obtained from the corresponding static black hole solution by performing a double Wick rotation. The validity of this approach was succesfully tested for static black hole solutions arising from different self-interactions of the minimally-coupled scalar field in \cite{Correa2012} and for a rotating hairy black hole in \cite{Correa2013}. 

There exist various generalizations of the Cardy formula for warped AdS$_3$ black holes, Lifshitz black holes and hyper-scaling violation black holes (both in 3d and higher dimensions) where it is also possible to incorporate rotation and electric charge. The reader is referred to \cite{Donnay:2015iia,Detournay:2016gao,Gonzalez:2011nz,Shaghoulian:2015dwa,Bravo-Gaete:2015wua,Bravo-Gaete:2015iwa,Ayon-Beato:2015jga,Verlinde2000,Cai2001,Shaghoulian:2015kta,Shaghoulian:2015lcn,Ayon-Beato:2019kmz,BravoGaete:2017dso} for details. Later, it was realized that the semi-classical entropy of the Schwarzschild black hole (with a spherical horizon) in $d$-dimensions might be reproduced by a proper modification of it \cite{Hassaine:2019uyn}.

For 3d static black holes, which we will be interested in here, one has $\tD^\pm_0 = \frac{1}{2} M_0 \ell$, where $M_0$ is the mass of the soliton, and the Cardy formula \eqref{Cardynew} simplifies into
\begin{equation}
	\SC = 4 \pi \ell \sqrt{-M_0 M}.\label{Cardyst}
\end{equation}
If the mass $M$ and the semi-classical entropy $S$ of the black hole are known, it suffices to determine the mass of the corresponding soliton solution $M_0$ to check whether the Cardy formula \eqref{Cardyst} yields a microscopic derivation of the entropy. 

The soliton configuration corresponding to the static BTZ black hole with the line element \eqref{BTZgr} obtained by double Wick rotation
\begin{equation}
	t_b \to i \ell \th_s, \qquad \qquad \qquad \th_b \to i \frac{t_s}{\ell},\label{wick}
\end{equation}
is just the AdS$_3$ soliton \cite{Horowitz:1998ha} whose line element reads
\begin{equation}
	\dd{s}^2_s = -\frac{r^2}{\ell^2} \dd{t}^2_s + \frac{\ell^2}{r^2-\ell^2} \dd{r}^2 + (r^2 - \ell^2) \dd{\th}^2_s.\label{AdSsol}
\end{equation}
By a redefinition of the radial coordinate $r^2 - \ell^2 \to r^2$, it becomes the line element of the AdS$_3$ spacetime in global coordinates
\begin{equation}
	\dd{s}^2 = -\left[1+\frac{r^2}{\ell^2}\right] \dd{t}^2 + \left[1+\frac{r^2}{\ell^2}\right]^{-1} \dd{r}^2 + r^2\dd{\th}^2,
\end{equation}
which shows that the ground state for the BTZ black hole is the global AdS$_3$ spacetime. The mass of the AdS$_3$ soliton in \eqref{AdSsol} can be easily computed to be\footnote{See \cite{Correa2011} for a Hamiltonian derivation.}
\begin{equation}
	M_0 = -\frac{1}{8 G},\label{M0AdS}
\end{equation}
which reproduces the semi-classical entropy given in \eqref{BTZth} when used in Cardy formula \eqref{Cardyst}.

In order to understand why Strominger's derivation works, one can check the Cardy formula with the central charges \eqref{Cardystan} for static black holes
\begin{equation}
	\SC = 2 \pi \sqrt{\frac{c^\pm}{3} M \ell}.\label{CardycGR}
\end{equation}
Since the mass of the AdS$_3$ soliton \eqref{M0AdS} can be expressed in terms of the central charges of GR \eqref{cGR} as
\begin{equation}
	M_0 = - \frac{c^\pm}{12 \ell},\label{M0toc}
\end{equation}
two versions of Cardy formula (\ref{CardycGR}, \ref{Cardyst}) become equivalent\footnote{This also holds when the rotation is turned on ($J \neq 0$).}. However, the soliton mass $M_0$ and the central charges $c^\pm$ of a generic theory are generally not related in the same manner. This shows why Strominger's derivation works for the BTZ black hole in GR, and also justifies the need to use the Cardy formula without central charges \eqref{Cardyst} in more complicated cases.

In this paper, we aim to provide further non-trivial evidence for the role played by the soliton solution in the microscopic derivation of the entropy of hairy black holes. For this purpose, we will focus on certain examples of 3d Horndeski theories, which are scalar-tensor theories with 2nd-order field equations \cite{Horndeski1974,Kobayashi2019}.  They were recently obtained from 2nd- and 3rd-order Lovelock Lagrangians, which normally give non-trivial contributions to field equations only in $d>4$ and $d>6$ respectively, via various methods (see \cite{Fernandes2020,Kobayashi2020,Lu2020,Hennigar2020,Alkac2022}). Lovelock gravities are the most natural generalizations of GR that are pure gravity theories with 2nd-order field equations \cite{Lovelock:1971yv,Lovelock:1972vz} (see \cite{Padmanabhan2013} for a review)  admitting exact static black hole solutions \cite{Wheeler1986, Wheeler1986a}. After a regularized Kaluza-Klein reduction or a technique called the ``Weyl trick'', one obtains Horndeski theories admitting hairy black hole solutions in $d=3$. We will call them 3d Lovelock gravities because of their higher-dimensional origin.

In previous works, either the Regge-Teitelboim method \cite{Regge1974} (by using the full form of the Hamiltonian of the theory under consideration) or the quasi-local generalization of the Abbot-Deser-Tekin method  \cite{Abbott:1981ff,Deser:2002rt,Deser:2002jk} given in \cite{Kim:2013zha,Gim:2014nba} have been used to find the soliton mass. However, both become quite cumbersome for 3d Lovelock gravities\footnote{See \cite{Peng:2015yjx} for the quasi-local charges of Horndeski theories.}. By considering a mini-superspace of static field configurations in the Regge-Teitelboim method, which can also be used to study the thermodynamics of the black hole solutions, we will be able to obtain the soliton mass and verify that it has zero entropy by an appropriate choice of mini-superspace variables. We would like to emphasize that this is the first time the whole analysis is performed via the mini-superspace formalism, and therefore, our work shows that even more complicated scenarios such as higher-dimensional planar black holes with rotation and charge can be studied similary.

\section{3d Einstein-Gauss-Bonnet theory}
\label{sec:3degb}
We start our analysis with 3d Einstein-Gauss-Bonnet (EGB) theory whose action is given by
\begin{align}
	I &= \frac{1}{16 \pi G} \int \dd^3{x} \sqdet  \left[R + \frac{2}{L^2} + \a_2 L^2 \cL_2 \right], \label{egb}\\
	\cL_2 &= 4 G^{\mu\nu}\phi_{\mu}\phi_{\nu} - 4 X \Box \phi + 2 X^2,
\end{align}
where $\a_2$ is a dimensionless coupling constant, $L$ is a constant with dimension of a length and $\cL_2$ is the 3d version of the well-known 2nd-order Lovelock Lagrangian, also known as the Gauss-Bonnet invariant, which reads
\begin{equation}
	\cG = R^2 - 4 R^{\m\n} R_{\m\n} + R^{\m\n\r\s} R_{\m\n\r\s}.
\end{equation}
For the terms with the scalar field $\f$, we use the following definitions
\begin{equation}
	\f_\m \equiv \pd_\m \f, \qquad \f_{\m\n} \equiv \nab_\m \nab_\n \f, \qquad  \square \f   \equiv g^{\m\n} \f_{\m\n}, \qquad X \equiv g^{\m\n} \f_\m \f_\n.\label{phidef}
\end{equation}
In order to study static field configurations, which cover both the black hole and the soliton solutions, we assume a line element of the form
\begin{equation}
	\dd s^2 = - \cN^2 \frac{r^2}{L^2} \cF  \dd{t}^2 + \frac{L^2}{r^2 \cF} \dd{r}^2 + \cR^2 \dd{\theta}^2,\label{minans}
\end{equation}
where the functions $(\cN, \cF, \cR)$ characterizing the spacetime metric, together with the scalar field $\f$, depend only the radial coordinate $r$. By inserting them directly into the action \eqref{egb}, one can obtain a reduced action for the functions $(\cN, \cF, \cR, \f)$ and obtain the equations satisfied by them from the Euler-Lagrange equations of the resulting reduced action. 

In order to study the thermodynamics of the solutions, one can identify the partition function for a thermodynamic ensemble with the Euclidean path integral evaluated by the saddle-point approximation around the Euclidean continuation of the classical solution \cite{Gibbons1977}. Assuming $\b$ and $\eta$ are the periodicities of the Euclidean time $\t = i t$ and the angular coordinate $\th$ respectively, the Euclidean reduced action can be put into the following form
\begin{align}
	\IE =&\,\, - \frac{\eta \beta }{16\pi G L^2}\int \dd{r} \cN \Bigg[ 2 \cR - r \bigg[ r \cF^{\prime} \cR^{\prime} + 2 \cF (\cR^{\prime} + r \cR^{\prime \prime}) \bigg]  \nonumber\\
	& +2 \alpha_2 r^3 \xi \cF \bigg[ r \xi^3 \cF \cR +2 \xi^2 \cR (2\cF + r \cF^{\prime}) - 4r \cF \xi^{\prime} \cR^{\prime} \nonumber\\
	&  + \xi \Big( -3r \cF^{\prime} \cR^{\prime} + \cF (4r\cR\xi^{\prime} - 6 \cR^{\prime} - 2r\cR^{\prime\prime}) \Big)  \bigg] \Bigg]+B,\label{egbred}
\end{align}
 where $\xi = \f^\pr$ and  $B$ is a boundary term. This form of the Euclidean reduced action is obtained by performing integration by parts successively until there remains no term with any derivative of the function $\cN$. The boundary term $B$ is fixed by requiring that the Euclidean action $\IE$ has an extremum on-shell\footnote{Thanks to this condition, the functional derivatives of the constraints in the Hamiltonian formulation of the theory are well-defined, and as a result, the symmetries are generated properly\cite{Regge1974}.}, i.e., $\var \IE\onsh =0$. In our case, this condition gives the variation of the boundary term $B$ as
\begin{align}
	\delta B = \frac{\eta\beta}{16\pi GL^2} \bigg[ &r \left( 1+2\alpha_2r^2\xi^2\cF \right)\left( r\cN\cF^{\prime}+2\cF (\cN+r\cN^{\prime}) \right)\delta\cR -2r^2\cF\cN \left( 1+2\alpha_2r^2\xi^2\cF \right)\delta\cR^{\prime}  \nonumber\\
	& - r^2\cN \left( \cR^{\prime} - 4\alpha_2r^2\xi^3\cF\cR + 6\alpha_2r^2\xi^2\cF\cR^{\prime} \right) \delta\cF + 8\alpha_2r^4\xi\cF^2\cN \left( \xi\cR-\cR^{\prime} \right)\delta\xi\bigg].\label{egbbou}
\end{align}
Note that the Euler-Lagrange equation following from the variation of the reduced action $\IE$ with respect to the function $\cN$, which is  $\fdv{\IE}{\cN}=0$, guarantees that there will be no bulk contribution to the on-shell Euclidean action. Therefore, the on-shell action can be found from the boundary term as
\begin{equation}
	\IE\onsh = B(\infty) - B(\rmin),\label{onsh}
\end{equation}
where $\rmin$ is the minimum value of the radial coordinate $r$. Since it is related to the free energy as $\IE\onsh = \beta F = \beta M - S$, the mass $M$ and the semi-classical entropy $S$ of the solution can be calculated from the following relations
\begin{equation}
	M = \partial_\beta \IE\onsh, \qquad \qquad  S = \left(\beta \partial_\beta - 1\right)\IE\onsh.\label{ther}
\end{equation}
\subsection{Black hole solution given in  [46]}
An ansatz for the line element of a static black hole solution can be written as 
\begin{equation}
	\dd{s}^2_b = -N^2 \frac{r^2}{L^2} f_b \dd{t}^2_b + \frac{L^2}{r^2 f_b} \dd{r}^2+r^2 \dd{\theta}^2_b,\label{bhans}
\end{equation}
which corresponds to the following choice of the mini-superspace variables
\begin{equation}
	\cN = N, \qquad  \qquad \cF = f_b, \qquad \qquad \cR = r.\label{egbcho}
\end{equation}
When $N=1$, which will be the case for the solutions that we study here, it describes an asymptotically AdS$_3$ black hole solution provided that the function $f_b$ has at least one zero and its value at infinity is a positive constant [$f_\infty \equiv f_b(r \to \infty) >0$]. The AdS radius in such a case is given by
\begin{equation}
	\ell = \frac{L}{\sqrt{\finf}}.\label{ldef}
\end{equation}
The periodicity $\b$ of the Euclidean time $\t_b = i t_b$, which is the inverse temperature of the black hole, is found by avoiding a conical singularity at the horizon. We can take the periodicity of the angular coordinate $\eta$ to be $2\pi$. Defining
\begin{equation}
	h_b \equiv \frac{r^2}{L^2} f_b,\label{hbtofb}
\end{equation}
we have
\begin{equation}
\label{eq:general_beta}
	\b = \frac{4 \pi}{h_b^\pr(\rp)} = \frac{4 \pi L^2}{r_+^2 f^\pr_b(r_+)}, \qquad \qquad  \eta = 2\pi, 
\end{equation}
where $r_+$ is the largest positive root of the equation $f_b(r)=0$ (or equivalently $h_b(r)=0$).

By using the choices \eqref{egbcho} in the reduced action of 3d EGB gravity \eqref{egbred}, the following field configuration can be deduced \cite{Hennigar2020a}
\begin{equation}
	N = 1,\qquad \f = \log(\frac{r}{\ell}),\qquad 1-f_b-\a_2 f_b^2 = \left(\frac{r_+}{r}\right)^2.\label{egbnov}
\end{equation}
The value of the function $f_b$ at infinity $\finf$ can be found by solving the equation
\begin{equation}\label{egbvac}
	1-\finf- \a_2 \finf^2=0,
\end{equation}
from which the AdS radius $\ell$ can be determined by using \eqref{ldef}. This solution is the analogue of  the higher-dimensional black hole solution of the EGB theory that exist in $d>4$ \cite{Boulware1985}.

The thermodynamic analysis of the black hole solutions that we consider here via the Euclidean methods was given in \cite{Alkac:2023mvr} but we will review the points that are relevant for the analysis of the corresponding soliton solutions. As discussed in \cite{Alkac:2023mvr}, the quadratic equation for the function $f_b$ can be solved. However, its explicit form is not needed for the thermodynamic analysis. This will be particularly useful when we study the black hole solution in 3d cubic Lovelock gravity where the function $f_b$ satisfies a cubic equation. By using $f_b(r_+)=0$ in the derivative of the equation satisfied by the function $f_b$ \eqref{egbnov}, one gets $f_b^\prime(r_+) = \frac{2}{r_+}$. From this, the inverse temperature $\b$ of the black hole follows as
\begin{equation}
	\b = \frac{2 \pi L^2}{r_+}.\label{temp}
\end{equation}
When the solution for the scalar field in \eqref{egbnov} is used as $\xi = \frac{1}{r}$, the variation of the boundary term $B_b$ simplifies into
	\begin{equation}
	\label{varBdyb}
	\var{B_b} = - \frac{\beta r^2}{8 GL^2}  \left( 1+2\alpha_2f_b \right)\delta f_b,
\end{equation}
where we have used $\eta = 2\pi$. 

The next step is to evalute it at infinity and the event horizon to find the value of the boundary term $B_b$ at these points, which will yield the on-shell value of the action through \eqref{onsh} ($\rmin = r_+$). The variation of the function $f_b$ at these values of the radial coordinate $r$ are as follows,
\begin{equation}
	\var{f_b}\eval_{\infty}= - \frac{2\rp}{ (1+2\a_2 f_b)r^2} \delta r_+\eval_{\infty},\qquad \qquad
	\var{f_b}\eval_{r_+}= -f_b^\pr(\rp) \var{\rp}=-\frac{4 \pi L^2}{\b r_+^2} \var{r_+}.\label{delfb}	
\end{equation}
The variation at any $r>\rp$ can be obtained from the equation satisfied by the function $f_b$ in \eqref{egbnov}, the first equation is just the variation evaluated at infinity. Note that it generates a finite contribution in $\delta B_b$, as it can be seen by replacing \eqref{delfb} in \eqref{varBdyb}. On the other hand, the second one requires a special care. The variation of the function $h_b$ at the horizon should respect the conditions $h_b(\rp)=0$ and $(h_b + \var{h_b})(\rp + \var{\rp}) = 0$, from which one gets
\begin{equation}
	\var{h_b}\eval_{r_+} = -h_b^\pr(\rp) \var{\rp} = -\frac{4 \pi}{\b} \var{\rp},
\end{equation}
where we have used \eqref{eq:general_beta}. Using this with the relation between the functions $h_b$ and $f_b$ \eqref{hbtofb} yields the result in \eqref{delfb}.

With these at hand, the on-shell action can be obtained easily as
\begin{equation}
	\IE^b\onsh = B_b(\infty) - B_b(\rp) =\frac{\b \rp^2}{8 G L^2} -  \frac{\pi \rp}{2 G},
\end{equation}
from which the mass $M$ and the entropy $S$ are found by using\footnote{See \cite{Hennigar2020a} for a derivation from the Wald formula.} \eqref{ther} as
\begin{equation}
	M = \frac{\rp^2}{8 \finf G \ell^2}, \qquad \qquad  S = \frac{\pi \rp}{2 G}.\label{thermegb}
\end{equation}
Here, we have used \eqref{ldef} to write the mass $M$ in terms of the AdS radius $\ell$. This completes our review of the thermodynamics of the  black hole solution.

The holographic studies of 3d Lovelock gravities was initiated in \cite{Alkac2023}, where it was shown that, unlike a generic Horndeski theory, it is possible to construct a monotonic c-function along a renormalization group flow induced by matter fields and to establish a holographic c-theorem along the lines of \cite{Myers:2010xs, Myers:2010tj}. The value of the c-function in the UV ($r \to \infty$) is proportional to the Weyl anomaly coefficient, from which the central charges for the 3d EGB theory can be read as\footnote{In \cite{Alkac2023}, $\ell=L$ is used for simplicity such that $\finf =1$.}
\begin{equation}\label{cegb}
	c^\pm= \frac{3\ell}{2 G} \left(1-2\a_2 \finf\right).
\end{equation}
This result was also confirmed by a direct calculation of the Weyl anomaly coefficient of the 2d boundary CFT by considering the theory on S$^2$. It is obvious that the Cardy formula with central charges for static black holes \eqref{CardycGR} fails to reproduce the semi-classical entropy in \eqref{thermegb}. Therefore, we need to construct the soliton solution describing the ground state.
\subsection{Soliton corresponding to the black hole solution}
As mentioned in the introduction, the soliton solution is obtained by applying a double Wick rotation, which we perform as
\begin{equation}
	t_b \to i L \th_s, \qquad \qquad \th_b \to i \frac{t_s}{L}.\label{wickegb}
\end{equation}
When applied to the ansatz \eqref{bhans} that we used to obtain the black hole solution, one gets
\begin{equation}
	\dd{s}^2_s = -\frac{r^2}{L^2} \dd{t}^2_s + \frac{L^2}{r^2 f_s} \dd{r}^2+N^2 r^2 f_s \dd{\theta}^2_s,\label{solans}
\end{equation}
The bulk piece of the reduced action remains the same after the Wick rotation and one obtains the following field configuration describing the soliton 
\begin{equation}
	N = 1,\qquad \f = \log(\frac{r}{\ell}),\qquad 1-f_s-\a_2 f_s^2 = \left(\frac{r_+}{r}\right)^2.\label{egbsol}
\end{equation}
The event horizon radius $\rp$ is technically an integration constant and it is known that the soliton solution describing the ground state should be devoid of integration constants \cite{Correa2011, Correa2012}. We will see that it will be fixed by matching the boundary geometries of both the black hole and the soliton solutions.

First of all, we should still demand the regularity of the spacetime at $r=\rp$. Denoting the periodicities of the Euclidean time $\t_s = i t_s$ and the angular coordinate $\th_s$ as $\b_s$ and $\eta_s$ respectively, and defining
\begin{equation}
	h_s \equiv \frac{r^2}{L^2} f_s,\label{hstofs}
\end{equation}
we see that it does not constrain $\b_s$ but we get
\begin{equation}
	\eta_s = \frac{4 \pi}{h_s^\pr(\rp)L} = \frac{4\pi L}{\rp^2 f_s^\pr(\rp)}.\label{defeta}
\end{equation}
For our solution, $f_s^\pr(\rp)=\frac{2}{\rp}$ and we obtain
\begin{equation}
	\eta_s = \frac{2 \pi L}{\rp}.\label{etas}
\end{equation}

The boundary geometries of both the black hole and the soliton solutions are the same provided that 
\begin{align}
\label{eq:matching_conditions}
	\left(\sqrt{-g_{00}}\, \b\right)_b &= \left(\sqrt{-g_{00}}\, \b\right)_s, \qquad \text{as } r\to\infty,\\
	\left(\sqrt{g_{33}}\, \b\right)_b &= \left(\sqrt{g_{33}}\, \b\right)_s, \qquad \,\,\,\,\, \text{as } r\to\infty,
\end{align}
which imply
\begin{equation}
	\b_s = \sqrt{\finf}\, \b, \qquad \qquad  \eta_s = \frac{2\pi}{\sqrt{\finf}}, \qquad  \qquad \eta_s \b_s = 2\pi \b.\label{bounmat}
\end{equation}
Using this value of the angular periodicity $\eta_s$ in \eqref{etas} fixes the integration constant $\rp$ as
\begin{equation}
	\rp = \sqrt{\finf}\, L \qquad \text{for the soliton.}\label{rpfixed}
\end{equation}
Since the field variations are taken with respect to the integration constants, one has to be careful in using this result while calculating the thermodynamical properties of the soliton. We can use \eqref{rpfixed} only after integrating the boundary term and obtaining the expressions for the mass $M_0$ and the entropy $S_0$ in terms of the integration constant $\rp$ from the on-shell action.

For the thermodynamical analysis, we should first note that the soliton ansatz \eqref{solans} corresponds to the following choice of the minisuperspace variables
\begin{equation}
	\cN = \frac{1}{\sqrt{f_s}},\qquad \qquad \cF = f_s, \qquad \qquad \cR=N r \sqrt{f_s}.\label{solcho}
\end{equation}
Using them together with $N=1$ and $\xi = \phi^\pr = \frac{1}{r}$ in \eqref{egbbou} gives the variation of the boundary term for the soliton as
\begin{equation}
	\var{B_s} = - \frac{\beta r^2}{8 GL^2}  \dv{r} \bigg[ r \Big( 1+2\alpha_2 f_s \Big) \delta f_s \bigg].
\end{equation}
The variation of the function $f_s$ at infinity is obtained from the equation satisfied by it in \eqref{egbsol} as follows
\begin{equation}
		\var{f_s}\eval_{\infty}= - \frac{2r_+}{ (1+2\a_2 f_s)r^2} \delta r_+\eval_{\infty}.
\end{equation}
At $r=\rp$, we make use of that the variation of the function $h_s$ should respect the conditions
\begin{equation}
	(h_s + \var{h_s}) (\rp + \var{\rp}) = 0, \qquad \qquad (h_s^\pr + \var{h_s^\pr}) (\rp + \var{\rp}) = \frac{4 \pi}{\eta_s L},
\end{equation}
which imply
\begin{equation}
	\var{h_s}\eval_{r_+} = - h_s^\pr(\rp) \var{\rp}, \qquad \qquad \var{h_s^\pr}\eval_{r_+} = - h_s^{\pr\pr}(\rp) \var{\rp}.
\end{equation}
Together with the relation \eqref{hstofs}, they yield the following relations for the function $f_s$
\begin{equation}
	\var{f_s}\eval_{r_+} = -  f_s^\pr(\rp) \var{\rp}, \qquad  \var{f_s^\pr}\eval_{r_+} = - \left[ f_s^{\pr\pr}(\rp) + \frac{2}{\rp} f^\pr(\rp)\right] \var{\rp}.\label{varfrp}
\end{equation}
We should keep in mind that the integration constant $\rp$ will be fixed according to \eqref{rpfixed} at the end. After the integration of the boundary term, one gets
\begin{equation}
	B_s(\infty) = - \frac{\rp^2}{8GL^2}, \qquad \qquad B_s(\rp) = 0,
\end{equation}
where we have used 
\begin{equation}
	f_s^\pr(\rp) = \frac{2}{\rp},\qquad \qquad f_s^{\pr\pr}(\rp) = - \frac{2(3+4\a_2)}{\rp^2},\label{fsders}
\end{equation}
for our solution.

Using the value of the integration constant from \eqref{rpfixed}, we obtain the on-shell action for the soliton as
\begin{equation}
	\IE^s\onsh = -\frac{\b \finf}{8G}.
\end{equation}
The mass $M_0$ and the entropy $S_0$ for the soliton follows from the relations \eqref{ther} easily as
\begin{equation}
	M_0 = -\frac{\finf}{8 G},\qquad \qquad S_0 = 0.\label{thermegbsol}
\end{equation}
The vanishing of the entropy verifies the non-degeneracy of the ground state. More importantly, using the mass of the soliton $M_0$ in the Cardy formula without central charges for static black holes \eqref{Cardyst} reproduces the semi-classical entropy that we have derived in \eqref{thermegb}. The formula with the central charges \eqref{CardycGR} is not succesful since the soliton mass is not related to the central charges as given in \eqref{M0toc}.

Note that the Cardy formula holds as long as the equation \eqref{egbvac} is satisfied with an $\finf>0$, which is a condition that follows from the definition of the AdS radius $\ell = \frac{L}{\sqrt{\finf}}$. This means that the Cardy formula does not directly restrict the sign of the central charge in \eqref{cegb} such that one can have a unitary ($c\geq0$) or non-unitary ($c<0$) boundary CFT depending on the value of $\finf$. The boundary theory is unitary for $\finf \geq \frac{2}{3}$ where the equality holds for the anomaly free theory ($c=0$) and non-unitary for $0<\finf<\frac{2}{3}$. It is always possible to obtain a value of the coupling $\a_2$ corresponding to a given $\finf$ from \eqref{egbvac}.

As in the case of Lovelock gravities in higher-dimensions ($d>4$), it is possible to make a choice for the coupling $\a_2$ such that one has a unique vacuum, i.e., a unique solution of \eqref{egbvac} for $\finf$. Traditionally, such a point is called a Chern-Simons point in odd dimensions  since the corresponding Lagrangian can be written as a Chern-Simons form obtained from the AdS curvature two-form (see \cite{Crisostomo2000} for details). Here, having a scalar-tensor theory, we lack such a description. At this point in the parameter space, one has
	\begin{align}
	&\finf = 2,\qq \a_2 = -\frac{1}{4}, \qq c^\pm = \frac{6 \ell}{2 G},\nn \\
	h_b = \frac{2 r^2}{L^2} &\left[1-\left(\frac{\rp}{r}\right)^2\right], \qq 	h_s = \frac{2 r^2}{L^2} \left[1-2\left(\frac{L}{r}\right)^2\right],\\
	M&= \frac{\rp^2}{16 G \ell^2},\qq S = \frac{\pi \rp}{2 G}, \qq M_0 = -\frac{1}{4G}.\nn
	\end{align}
This is a case where the metric function $h$ takes a particularly simple form and the boundary CFT is unitary.	
\section{3d Cubic Lovelock gravity}
Now, we would like to study a more complicated scenario where two distinct static black hole solutions exist. With this aim, we consider 3d cubic Lovelock gravity with the action \cite{Alkac2022}
\begin{align}
	I =&\,\, \frac{1}{16 \pi G} \int \dd^3{x} \sqdet \left[R + \frac{2}{L^2} + \a_2 L^2 \cL_2 +  \a_3 L^4 \cL_3 \right],\\
	\cL_2 = &\,\,4 G^{\mu\nu}\phi_{\mu}\phi_{\nu} - 4 X \Box \phi + 2 X^2,\\
	\cL_3=&\,-48R^{\mu\nu}\phi_{\mu\nu}X-48R^{\mu\nu}\phi_{\mu}\phi_{\nu}X+24 RX\Box\phi+6RX^2+96\phi_{\mu\nu}\phi^{\mu}\phi^{\nu}\Box\phi\nn\\
			&+48\phi_{\mu\nu}\phi^{\mu\nu}\Box\phi-24\phi_{\mu\nu}\phi^{\mu\nu}X
			-144\phi_{\mu\nu}\phi^{\mu}\phi^{\nu}X-96\phi^{\mu}\phi^{\nu}\phi_{\mu\rho}\phi_{\ \nu}^{\rho}-32\phi^{\mu\nu}\phi_{\mu\rho}\phi_{\ \nu}^{\rho}\nn\\
			&-16(\Box\phi)^3+24X(\Box\phi)^2-24X^3,\label{cl}
\end{align}
where we have again employed the definitions \eqref{phidef}. $\cL_2$ is the 3d version of the Gauss-Bonnet invariant as discussed before and $\cL_3$ is the 3d version of the cubic Lovelock Lagrangian given by
\begin{equation}
	\cC = \frac{1}{8} \d^{\m_1\n_1\m_2\n_2\m_3\n_3}_{\r_1\s_1\r_2\s_2\r_3\s_3} R^{\r_1 \s_1}_{\m_1\n_1} R^{\r_2 \s_2}_{\m_2\n_2} R^{\r_3 \s_3}_{\m_3\n_3}.
\end{equation}
$\a_3$ is another dimensionless coupling constant just as $\a_2$.

The static black hole solutions of this theory were found in \cite{Alkac2022} and their thermodynamics were presented in \cite{Alkac:2023mvr} by finding the reduced action of the theory for the black hole ansatz \eqref{bhans} and $\f = \f(r)$. However, as we have seen in the previous section, one can start with the reduced action for our more general mini-superspace ansatz \eqref{minans}. Then, it is enough to make proper choices for the mini-superspace variables to derive both the black hole and the soliton solutions, and also their thermodynamics. After quite a lengthy calculation, one finds the Euclidean reduced action of 3d cubic Lovelock gravity as
\begin{align}
	\IE =&\,\, - \frac{\eta \beta }{16\pi G L^2}\int \dd{r} \cN \Bigg\{ 2 \cR - r \bigg[ r \cF^{\prime} \cR^{\prime} + 2 \cF (\cR^{\prime} + r \cR^{\prime \prime}) \bigg]  \nonumber\\
	& +2 \alpha_2 r^3 \xi \cF \bigg[ r \xi^3 \cF \cR +2 \xi^2 \cR (2\cF + r \cF^{\prime}) - 4r \cF \xi^{\prime} \cR^{\prime} \nonumber\\
	&  -\xi \Big( 3r \cF^{\prime} \cR^{\prime} - \cF (4r\cR\xi^{\prime} - 6 \cR^{\prime} - 2r\cR^{\prime\prime}) \Big)  \bigg] \nonumber\\
	&  -6\alpha_3 r^5\xi^3 \cF^2 \bigg[ 4r\xi^3 \cF \cR - 24r \cF \xi^{\prime} \cR^{\prime} + 12\xi^2\cR (2\cF+r\cF^{\prime}) \nonumber\\
	& -3\xi \Big(5r \cF^{\prime}\cR^{\prime}-2\cF(4r\cR\xi^{\prime} - 5\cR^{\prime} - r\cR^{\prime\prime})\Big)\bigg] \Bigg\}+B, \label{clred}
\end{align}
where the variation of the boundary term, which follows from the condition $\var \IE\onsh =0$, is given by
\begin{align}
	\delta B = \frac{\eta \beta  r}{16 \pi G L^2} \Bigg[  &\big( 1 + 2\alpha_2 r^2 \xi^2\cF - 18\alpha_3 r^4 \xi^4 \cF^2\big) \big(r\cN\cF^{\prime} + 2\cF (\cN+r\cN^{\prime}) \big) \delta \cR \nonumber\\
	& +2r\cF\cN  \Big( -1 -2\alpha_2 r^2 \xi^2 \cF + 18 \alpha_3 r^4 \xi^4 \cF^2 \Big) \delta \cR^{\prime} \nonumber\\
	& -r\cN  \Big( \cR^{\prime} -2\alpha_2 r^2 \xi^2 \cF (2\xi \cR - 3 \cR^{\prime}) +18 \alpha_3 r^4 \xi^4 \cF^2 (4\xi \cR - 5\cR^{\prime}) \Big)\delta \cF \nonumber\\
	& +8r^3\cN \xi \cF^2  \big(\alpha_2 - 18r^2\alpha_3 \xi^2 \cF \big)(\xi \cR - \cR^{\prime})\delta\xi \Bigg].\label{delbcubic}
\end{align}
\subsection{Black hole solution given in [33]}
3d cubic Lovelock gravity  admits a black hole solution analogous to the higher-dimensional black hole solution of cubic Lovelock gravity that exist in $d>6$ \cite{Clunan2004},  where the functions in our black hole ansatz \eqref{bhans} and the scalar field $\f$ takes the following form \cite{Alkac2022}
\begin{equation}
	N=1, \qquad 1 - f_b - \alpha_2 f_b^2 + 6\alpha_3 f_b^3 = \left(\frac{\rp}{r}\right)^2, \qquad \f = \log(\frac{r}{\ell}).
\end{equation}
This time, the value of the function $f_b$ at infinity $\finf$ satisfies the cubic equation
\begin{equation}\label{cubvac}
	1 - \finf - \alpha_2 \finf^2 + 6\alpha_3 \finf^3=0.
\end{equation}
The mass and the entropy of the solution can be obtained by taking the mini-superspace variables as given in \eqref{egbcho} in the variation of the boundary term \eqref{delbcubic}. The procedure was reviewed in the previous section and details can be found in  \cite{Alkac:2023mvr}. The mass $M$ and the entropy $S$ the black hole are found as 
\begin{equation}
	M= \frac{\rp^2}{8 \finf G \ell^2}, \qquad \qquad S = \frac{\pi\rp}{2G}.\label{thermcubic}
\end{equation}

The central charges for the theory was computed in \cite{Alkac2023} to be
\begin{equation}
	c^\pm = \frac{3\ell}{2 G} \left(1-2\a_2 \finf+6 \a_3 \finf^2\right).
\end{equation}
Using them in the formula \eqref{CardycGR} obviously does not yield the semi-classical entropy $S$ in \eqref{thermcubic}.
\subsection{Soliton corresponding to the black hole solution}
The corresponding soliton solution following from the double Wick rotation \eqref{wickegb} of the black hole solution is as follows 
\begin{equation}
	N=1, \qquad 1 - f_s - \alpha_2 f_s^2 + 6\alpha_3 f_s^3 = \left(\frac{\rp}{r}\right)^2, \qquad \f = \log(\frac{r}{\ell}), \qquad \rp = \sqrt{\finf} \, L,\label{cubsol}
\end{equation}
where we fix the integration constant $\rp$ by obeying the constraints imposed by the regularity of the spacetime \eqref{etas} and the boundary matching conditions \eqref{bounmat}. Note that we again have $f_s(\rp) = \frac{2}{\rp}$, which can be verified from the equation satisfied by the function $f_s$.  Using the choice \eqref{solcho} of the minisuperspace variables in \eqref{delbcubic} together with $N=1$ and $\f = \log(\frac{r}{\ell})$ from \eqref{cubsol}, we obtain the variation of the boundary term for the soliton as
\begin{equation}
	\delta B_s = - \frac{  \beta r^2}{8 G L^2} \dv{r} \Big[ r \Big( 1 + 2\alpha_2 f_s -18\alpha_3 f_s^2 \Big)\delta f_s \Big].
\end{equation}
While the variation of the function $f_s$ at infinity is given by
\begin{equation}
		\var{f_s}\eval_{\infty}= - \frac{2r_+}{ (1+2\a_2 f_s-18 \a_3 f_s^2)r^2} \delta r_+,
\end{equation}
the variation at $r=\rp$ can be easily calculated through relations \eqref{varfrp} where the equations \eqref{fsders} are still valid. As a result, we get
\begin{equation}
	M_0 = -\frac{\finf}{8 G},\qquad \qquad  S_0 = 0,
\end{equation}
for the soliton. 

Note that we exactly have the same structure as we had for the black hole solution in the 3d EGB theory [see \eqref{thermegb} and \eqref{thermegbsol}] and the semi-classical entropy is reproduced through the Cardy formula without the central charges \eqref{Cardyst} as long as provided that the equation \eqref{cubvac} has a positive root for $\finf$. For a given $\finf >0$, we have the following possibilities
\begin{align}
	&\text{Non-unitary boundary CFT ($c>0$):}\qq \,\,\,\a_2>\frac{2 \finf -1}{\finf^2}, \qq \a_3 = \frac{-1+\finf + \alpha_2 \finf^2}{6 \finf^3}, \nn\\
   & \text{Anomaly-free boundary CFT ($c=0$):}\qq \a_2=\frac{2 \finf -1}{\finf^2}, \qq \a_3 = \frac{3 \finf -2}{6 \finf^3},\\
   & \text{Unitary boundary CFT ($c>0$):}\qq \,\,\,\,\qq\a_2<\frac{2 \finf -1}{\finf^2}, \qq \a_3 = \frac{-1+\finf + \alpha_2 \finf^2}{6 \finf^3}.\nn
\end{align}

Again, there exists a choice for the couplings $\a_2$ and $\a_3$ that yield a unique solution of \eqref{cubvac} for $\finf$. The resulting theory has a unitary boundary CFT and admits a simple form of the metric function $h$. At this point in the parameter space, we have
\begin{align}
\finf = 3,&\qq \a_2 = -\frac{1}{3}, \qq \a_3 = -\frac{1}{162}, \qq c^\pm = \frac{4 \ell}{ G}, \nn \\
h_b = \frac{3 r^2}{L^2} &\left[1-\left(\frac{\rp}{r}\right)^{\nicefrac{2}{3}}\right], \qq 	h_s = \frac{3 r^2}{L^2} \left[1-3^{\nicefrac{1}{3}}\left(\frac{L}{r}\right)^{\nicefrac{2}{3}}\right],\\
M&= \frac{\rp^2}{24 G \ell^2},\qq S = \frac{\pi \rp}{2 G}, \qq M_0 = -\frac{3}{8G}.\nn
\end{align}
\subsection{BTZ black hole}
Contrary to the 3d EGB theory, the BTZ black hole  is admitted as a solution in 3d Cubic Lovelock gravity where the functions in our black hole ansatz \eqref{bhans} and the scalar field $\f$ takes the following form \cite{Alkac2022}
\begin{equation}
	N=1,\qquad f_b = \finf \left[1-\frac{\rp^2}{r^2}\right],\qquad \phi =\frac{1}{3} \sqrt{\frac{\alpha_2}{2\alpha_3 \finf}} \log\left[\frac{r - \sq{r^2 - \rp^2}}{\ell}\right],
\end{equation}
where the value of the function $f_b$ at infinity is related to the coupling constants $\a_2$ and $\a_3$ as
\begin{equation}\label{finfbtz}
	\finf =  \frac{\alpha_2^3 + 972  \alpha_3^2}{54 \alpha_3\left(\alpha_2^2 + 18 \alpha_3\right) }.
\end{equation}
Considering the effect of Horndeski couplings to the field equations as an energy-momentum tensor $T_{\m\n}$, we can write $G_{\m\n} - \frac{1}{L^2} g_{\m\n} = T_{\m\n}$. This is a solution where the energy-momentum tensor is proportional to the metric tensor, i.e., $T_{\m\n} = \frac{1}{\tilde{L}^2} g_{\m\n}$ where $\tilde{L}$ is a function of the couplings $\a_2$ and $\a_3$. As a result, one has $G_{\m\n} - \frac{1}{\ell^2} g_{\m\n} = 0$, where $\ell = \frac{L}{\finf}$ is the AdS radius corresponding to the effective cosmological constant $\Lambda_{\text{eff}} = -\frac{1}{\ell^2}$ of the theory while $L$ is the AdS radius corresponding to the bare cosmological constant $\Lambda_0 = -\frac{1}{L^2}$ of the theory introduced at the action level. When $\finf = 1$, one has $\Lambda_{\text{eff}} = \Lambda_0$ and $\ell = L$, giving rise to a stealth solution ($T_{\m\n}=0$ on-shell). Unlike the first example in 3d \cite{Ayon-Beato:2004nzi}, the scalar field is independent of time and depends solely on the radial coordinate. We will comment more in this particular case after a general discussion of the Cardy formula.

One might also wonder what happens when $\a_2^2 + 18 \a_3 = 0$, for which the solution above is ill-defined. Imposing this relation at the action level gives a degenerate solution where the function $f$ is left undetermined and the coupling constants are fixed as $\a_2 = -\frac{1}{3}$ and $\a_3 = -\frac{1}{162}$.


The mass and the entropy of the solution can also be obtained by taking the mini-superspace variables as given in \eqref{egbcho} in the variation of the boundary term \eqref{delbcubic}. Details can be found in  \cite{Alkac:2023mvr}. The mass and the entropy of the BTZ black hole in 3d cubic Lovelock gravity are found as 
\begin{equation}
	M =  \left[ 1 + \frac{\alpha_2^2}{18\alpha_3} \right] \frac{\rp^2}{8  G \ell^2},\qquad \qquad S =  \Bigg[ 1 + \frac{\alpha_2^2}{18\alpha_3} \Bigg] \frac{\pi \rp}{2 G} .\label{thermocubic}
\end{equation}
When we attempt to reproduce the semi-classical entropy $S$ in by using Cardy formula with central charges for static black holes \eqref{CardycGR}, it does not work as expected. Hence, it is needed to construct the soliton describing the ground state as before.
\subsection{Soliton corresponding to the BTZ black hole: the global AdS$_3$ spacetime}
After a double Wick rotation given in \eqref{wickegb}, we obtain the following configuration describing a soliton
\begin{equation}
	N=1,\qquad f_s = \finf \left[1-\frac{\rp^2}{r^2}\right],\qquad \phi =\frac{1}{3} \sqrt{\frac{\alpha_2}{2\alpha_3 \finf}} \log\left[\frac{r - \sq{r^2 - \rp^2}}{\ell}\right],\label{btzcubic} 
\end{equation}
As explained in the introduction, this coincides with the global AdS$_3$ spacetime with AdS radius $\ell = \frac{L}{\sqrt{\finf}}$. This time, we have $f_s^\pr(\rp) = \frac{2 \finf}{r_+}$, and consequently the angular periodicity $\eta_s$ becomes
\begin{equation}
	\eta_s = \frac{2 \pi L}{\finf r_+}.
\end{equation}
With this, the boundary matching condition in  \eqref{bounmat} fixes the integration constant $\rp$ as
\begin{equation}
	\rp = \frac{L}{\sqrt{\finf}} = \ell\qquad \text{for the soliton,}\label{rpbtz}
\end{equation}
which should be imposed only after integrating the boundary term. Using the choice \eqref{solcho} in \eqref{delbcubic} together with $\xi = \f^\pr= \frac{1}{3} \sqrt{\frac{\alpha_2}{2\alpha_3}} \frac{1}{r \sqrt{f_s}}$ from \eqref{btzcubic} , we obtain the variation of the boundary term for the soliton as
\begin{equation}
	\var{B_s} = -\frac{\b r^2}{144G L^2 } \left[18+\frac{\a_2^2}{\a_3}\right] \dv{r} \left(r \var{f_s}\right).\label{delBsbtz}
\end{equation}
The variation of the function $f_s$ at infinity is as follows
\begin{equation}
	\var{f_s}\eval_{\infty} = - \frac{2 \finf \rp}{r^2} \d \rp.
\end{equation}
At the horizon, we use the relations \eqref{varfrp} with $f_s^\pr(\rp) = \frac{2 \finf}{\rp}$ and $f_s^{\pr\pr}(\rp) = -\frac{6 \finf}{\rp^2}$. The on-shell Euclidean action for the soliton $\IE^s$ can be easily found by integrating the variation of the boundary term in \eqref{delBsbtz}. After imposing \eqref{rpbtz}, the relations \eqref{ther} give the mass and the entropy of the soliton as 
\begin{equation}
	\label{soliton_mass_btz_like}M_0 = - \frac{1}{8G} \left(1+\frac{\alpha_2^2}{18 \alpha_3}\right) , \qquad \qquad S_0=0.
\end{equation}
Using the soliton mass $M_0$ in the Cardy formula for static black holes with no reference to central charges \eqref{Cardyst} reproduces the semi-classical entropy in \eqref{thermocubic}. For generic values of the coupling constants $\a_2$ and $\a_3$, one has a BTZ black hole solution with an AdS radius $\ell = \frac{L}{\finf}$ with $\finf$ given in \eqref{finfbtz}.

In this case, it is quite cumbersome to study the unitarity of the boundary CFT for any given $\finf >0$ since it dependence on the coupling constants $\a_2$ and $\a_3$ is quite complicated as can be seen from \eqref{finfbtz}. As an example, we focus on the stealth configuration with the vanishing energy momentum tensor ($\finf = 1$ and $\ell = L$), for which we have
\begin{align}
&\text{Non-unitary boundary CFT ($c>0$):}\qq \,\,\,\a_2>\frac{9}{17}, \qq \a_3 = \frac{\a_2}{54}, \nn\\
& \text{Anomaly-free boundary CFT ($c=0$):}\qq \a_2=\frac{9}{17}, \qq \a_3 = \frac{\a_2}{54},\\
& \text{Unitary boundary CFT ($c>0$):}\qq \,\,\,\,\qq\a_2<\frac{9}{17}, \qq \a_3 = \frac{\a_2}{54},.\nn
\end{align}
This range of the parameters also ensures that the scalar field is well-defined ($\frac{\a_2}{\a_3}>0$) and the mass and the entropy of the black hole solution is positive ($1+\frac{\a_2^2}{18 \a_3}>0$). Furthermore, although the energy-momentum tensor vanish on-shell, the thermodynamic properties (including the ground state) are modified since $\a_2 \neq 0$ is required for a non-vanishing scalar field (see \cite{Bakopoulos:2024zke,Erices:2024iah} for recent examples of this phenomenon in the literature).


We have checked the validity of the Cardy formula in the context of 3d cubic Lovelock gravity. This theory is interesting because there are two different black hole solutions, with non-trivial scalar profiles. By means of the double Wick rotation, we have found two different ground states, and the Cardy formula holds for each one of them. This is not surprising, because the black hole solutions belong to disconnected sectors of the theory and, in the same way as in \cite{Correa2011}, there is no continuous way to deform one sector to the other, so each branch must possess its own ground state.

\section{Summary and outlook}
We have revisited the proposal of \cite{Correa2011} for the microscopic derivation of the entropy of 3d black holes where the semi-classical entropy is reproduced from the Cardy formula without central charges \eqref{Cardynew} describing the asymptotic growth of the number of states. In this setup, the ground state is described by a soliton obtained through a double Wick rotation of the static black hole solution. We have checked the validity of the formula \eqref{Cardyst} for static black holes in 3d Lovelock gravities, which are 3d Horndeski theories that were recently obtained by various procedures from the higher-dimensional Lovelock Lagrangians. In order to deal with the relatively complicated structure of these theories compared to previous studies, we employed the mini-superspace formalism to calculate the mass of the soliton required in the Cardy formula \eqref{Cardyst}. For all solutions we have considered, we have shown the validity of the proposal of \cite{Correa2011}, providing evidence in 3d Horndeski theories for the first time.

The natural way to extend our results further is to add other global charges to the static solutions studied here. It is known that rotating versions of static black hole solutions in 3d can be obtained by ``a boost in the $t-\th$ plane'' \cite{Lemos:1994xp,Lemos:1995cm} as follows: 
\begin{equation}
	t \to\frac{1}{\sqrt{1-\omega^2}}(t+\omega \ell \th), \qquad \qquad  \th \to \frac{1}{\sqrt{1-\omega^2}}(\th+\omega \frac{t}{\ell}).
\end{equation} 
For the  rotating black holes, the Cardy formula without central charges \eqref{Cardynew} reads
\begin{equation}
	\SC = 2 \pi \sqrt{-M_0 \ell (M \ell + J)} + 2 \pi \sqrt{-M_0 \ell (M \ell - J)},
\end{equation}
which we expect to hold for the rotating black holes resulting from the boost, as it was the case in \cite{Correa2013}.

Adding electric charge is another possibility, for which the soliton describing the ground state is magnetically charged and the Cardy formula is modified accordingly. The Cardy formula for charged AdS$_3$ black holes proposed in \cite{Bravo-Gaete:2015iwa} reads
\begin{equation}
	S_{\text{Cardy}}  = 4 \pi \ell \sqrt{\mid-M_0 +\a\, \Phi_m Q_m\mid\mid M-\a\, \Phi_e Q_e\mid},\label{chargedcardy}
\end{equation}
where $\left(\Phi_{e,m}, Q_{e,m}\right)$ are the electric/magnetic potentials and charges, and $\a$ is a constant that depends on the nature of electromagnetic coupling. Remarkably, the Smarr relation is intimately related to the Cardy formula (see the discussion in \cite{Bravo-Gaete:2015iwa}). Since the Smarr relation does not exist in 3d Eintein-Maxwell theory due to the logarithmic behaviour of the electric potential, the authors used a conformal coupling to electromagnetism to test the validity of this formula. The same strategy can be used for black holes in 3d Lovelock gravities. Later, it was shown that the mass and the electric charge of the black hole solution in Einstein-Maxwell theory get modified when the  ``holographic boundary conditions'' \cite{Perez:2015kea} are imposed such that the Smarr relation holds \cite{Erices:2017nta}. Therefore, one might expect the charged Cardy formula \eqref{chargedcardy} to be valid even with the usual Maxwell coupling when these boundary conditions are imposed. 

\paragraph*{Acknowledgements} G. A. is supported by T\"{U}B\.{I}TAK Grant No 124F058. L.G. thank Instituto de Matem\'aticas, Universidad de Talca, for its hospitality during the preparation of this work.

\bibliographystyle{utphys}
\bibliography{ref}

\end{document}